\documentclass[%
 amsmath,amssymb,
 aps,
prl,
floatfix,
]{revtex4-1}
\usepackage{geometry} 
\usepackage{graphicx}
\usepackage{epstopdf}
\usepackage[sort&compress]{natbib}

\begin{document}

\title{X-ray phase-contrast radiography and tomography with a multi-aperture analyser}
\author{M. Endrizzi$^1$}
\email{m.endrizzi@ucl.ac.uk}
\author{F.A. Vittoria$^1$}
\author{L. Rigon$^{2,3}$}
\author{D. Dreossi$^4$}
\author{A. Olivo$^1$}
\affiliation{$^1$ Department of Medical Physics and Biomedical Engineering, University College London, Gower Street, London WC1E 6BT, United Kingdom}
\affiliation{$^2$ Physics Department, University of Trieste, Via Valerio 2, 34127 Trieste, Italy}

\affiliation{$^3$ Istituto Nazionale di Fisica Nulceare, Sezione di Trieste, Via Valerio 2, 34127 Trieste, Italy}

\affiliation{$^4$ Sincrotrone Trieste SCpA, S.S. 14 km 163.5, 34012 Basovizza Trieste, Italy}

\begin{abstract}
We present a multi-aperture analyser set-up for performing X-ray phase contrast imaging in planar and three-dimensional modalities. A multi-slice representation of the sample is used to establish a quantitative relation between projection images and the corresponding three-dimensional distributions, leading to successful tomographic reconstruction. Sample absorption, phase and scattering are retrieved from the measurement of five intensity projections. The method is experimentally tested on custom built phantoms with synchrotron radiation: sample absorption and phase can be reliably retrieved also in combination with strong scatterers, simultaneously offering high sensitivity and dynamic range.
\end{abstract}

\maketitle

Conventional radiography is based on the attenuation of X-rays traversing a sample for generating contrast. In X-ray phase contrast imaging (XPCI) \cite{Bravin:13_PMB,Wilkins:14_PhilTransRSocA} additional contrast mechanisms, generating from the phase-shifts imparted to the beam by the sample, contribute to the image formation process. This can improve the visibility of a large variety of details of interest and its applications span across many different fields, encompassing material science, security screening, biology and medicine.
Amongst various implementations with which it is possible to obtain X-ray phase-contrast images \cite{Bonse:65a,Goetz:79,Davis:95,Ingal:95_JPhysD,
Wilkins:96,Chapman:97_PMB,Clauser:98_PATENT,
David:02_APL,Momose:03_JJAP,Mayo:04_OL,
Pfeiffer:06_NatPhys,Wen:09_Radiology,Morgan:12_APL,Miao:16_Nature}, we focus here on edge illumination (EI) \cite{Olivo:01_MedPhys}. EI can provide quantitative attenuation, phase \cite{Munro:12_PNAS} and dark-field \cite{Endrizzi:14_APL} representation of a sample; and can be adapted for use with synchrotron radiation, microfocus tubes and conventional rotating anode sources with extended focal spots \cite{Olivo:07_APL,Endrizzi:14_OL}. It exhibits negligible requirements in terms of temporal or spatial coherence \cite{Munro:10_OEb,Endrizzi:15_OE}, provides high sensitivity \cite{Marenzana:12_PMB,Diemoz:13_PRL,Diemoz:14_APL}, it is robust against mechanical and thermal instabilities \cite{Millard:13_RSI,Endrizzi:15_APL} and enables low-dose implementations of XPCI in planar and three-dimensional imaging \cite{Olivo:13_MedPhys,Hagen:14_MedPhys}.

We introduce here a new multi-aperture analyser set-up for performing edge illumination with synchrotron radiation in situations where very wide angular ranges must be explored, for example in the presence of strong scatterers. Through experimental tests on custom built phantoms, we show that refraction can be accurately retrieved independently from the presence of large amount of scattering. Two spatial resolutions are investigated, showing that the retrieved signals are independent from this parameter. A simple model based on a multi-slice representation of the sample enables the reconstruction of the three-dimensional images. The method retrieves absorption, refraction and dark-field images, simultaneously providing high sensitivity and large dynamic range for both planar and three dimensional XPCI applications.

\begin{figure*}
\includegraphics[width = \textwidth]{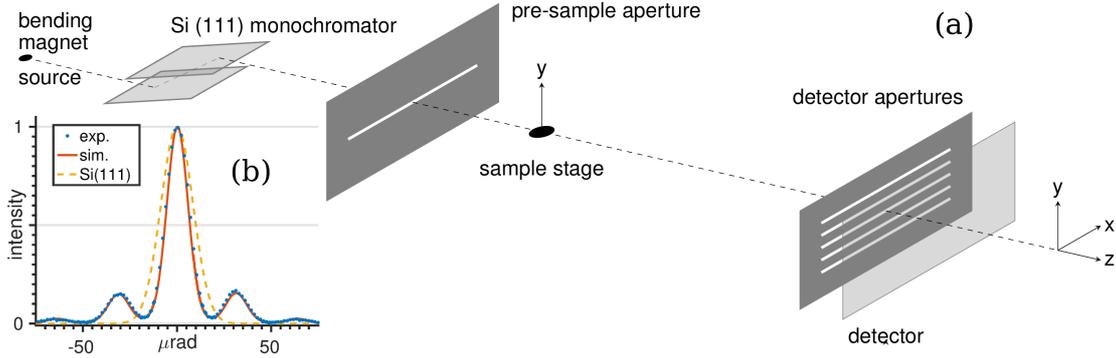}
\caption{\label{fig:1}Sketch of the experimental set-up: (a) monochromatic synchrotron radiation is shaped to a narrow laminar beam by an aperture positioned before the sample. It is then analysed by a set of apertures before detection. Two dimensional images are built by scanning the sample along the $y$ direction. (b) simulated (solid line) and experimental (circles) illumination functions. An approximate Si (111) rocking curve is also plotted (dashed line) for comparison (20 keV).}
\end{figure*}

The experimental set-up is sketched in Figure \ref{fig:1}(a). 
The beam from a bending magnet source is monochromatised with a double Si (111) crystal and then shaped by a narrow aperture, before it impinges on the sample. It is then analysed by a set of apertures positioned immediately before a digital detector. In order to achieve different levels of illumination, the detector and the detector apertures are scanned along the $y$ direction. The data collected with each detector exposure result in a single line of pixels, and a two-dimensional image is built by scanning the sample along the $y$ direction and exposing the detector multiple times.

The illumination function describes how the detected intensity varies depending on the relative displacement $\bar y$ between pre-sample and detector apertures. By using a geometrical optics model, the intensity measured at the detector is described as a convolution between the illumination function $L(\bar{y})$ and the object function $O(x,y)$:
\begin{eqnarray}\label{eq:conv_single}
I(\bar y) & = & \int L(\bar{y}-y) O(y) dy.
\end{eqnarray}
Transmission, refraction and scattering properties of the sample were retrieved by using a multi-Gaussian model for the illumination function \cite{Endrizzi:14_APL}:
\begin{eqnarray}\label{eq:mulit_gauss_model}
I(\bar y)  & = & \sum_{m}\sum_{n} A_{mn} \exp{\left[-\frac{{( \bar y-\mu_{mn})}^{2}}{2\sigma_{mn}^2} \right]}
\end{eqnarray}
where $L(\bar y) = $ $\sum_n  (A_n/\sqrt{2\pi\sigma_n^2})$ $\exp{[-{(\bar y -\mu_n)}^{2}/2\sigma_n^2]}$, ($n=1 \ldots N$); and  $O(\bar y)  =$ $\sum_m (A_m/\sqrt{2\pi\sigma_m^2})$ $\exp{[-{(\bar y-\mu_m)}^{2}/2\sigma_m^2]}$, ($m=1 \ldots M$). The parameters are defined as follows: $\mu_{mn} = \mu_m + \mu_n$, $\sigma_{mn}^2 = \sigma_m^2+\sigma_n^2$ and $A_{mn} = A_m A_n (1/\sqrt{2\pi\sigma_{mn}^2})$. We note that beam absorption and refraction can be both included in the object function $O(y)$ as a multiplicative factor $t$ and a shift $\Delta y_R$ of the center of the distribution, respectively. In this formulation, a purely absorbing object is represented by a Dirac's delta function centred in zero and multiplied by a factor $t$, which indicates the transmitted to incident intensity ratio. If the sample is also refracting, the delta function is laterally shifted $O(y) = t\delta(y-\Delta y_R)$, thus the effect on the illumination function is a reduction in the total transmitted intensity by the factor $t$ plus a $\Delta y_R$ lateral shift due to refraction. In the case of a sample exhibiting also dark-field contrast, the object function $O(y)$ has a finite width and the presence of the sample results into a broadened illumination function.
 
Phase images can be obtained by numerical integration of the refraction images by taking into account that the refraction angle $\alpha = (\lambda/2\pi) \partial_y \Phi(x,y)$ is directly proportional to the gradient of the object's phase shift $\Phi(x,y)$, where $\lambda$ is the wavelength. In the case of a parallel beam geometry, of which a synchrotron set-up is usually a good approximation, the refraction angle can be measured as the ratio between the relative shift of the illumination function and the sample-to-detector distance $\alpha = \Delta y_R / z_{sd}$.

A thick sample can be represented with a multi-slice approach, by separately considering the subsequent effect of thin sample sections along the beam axis $z$. We define $O_k$ a thin but finite thickness $\Delta_z$, section of the object along $z$. The entire object is obtained by the sum of its sections (with $k = 1 \ldots K$) and its extent along $z$ is given by $Z_{o} = \Delta_z K$. The intensity measured at the detector can then be expressed in terms of the contributions from separate layers
\begin{eqnarray}\label{eq:conv_multi_layer}
I(\bar y) & = & (L * O)(\bar y) \\
 & = & (L * O_1 * O_2 *  \ldots * O_K )(\bar y)
\end{eqnarray}
where the more compact notation $*$ was introduced to indicate convolution. Let's consider, for simplicity's sake, a single-Gaussian model for the object's section functions $O_k$. In this case, the entire object function $O(y) = (O_1 * O_2 *  \ldots * O_K)(y)$ is still a Gaussian, with mean $\mu_O = \sum_1^K \mu_k$, variance $\sigma_O^2 = \sum_1^K \sigma_k^2$ and amplitude $A_O = (1/\sqrt{2 \pi \sigma_O^2})\prod_1^K A_k$; this is now expressed as a function of the contributions from the individual thin layers. The intensity transmitted by the entire object becomes
\begin{eqnarray}\label{eq:att_tot_multislice}
t & = & \prod_1^K A_k \\
 & = & \prod_1^K e^{-(4 \pi \beta_k \Delta_z/\lambda)} 
\end{eqnarray}
and a similar result is obtained for the refraction
\begin{eqnarray}\label{eq:ref_tot_multislice}
\Delta y_R & = &\sum_1^K \mu_k \\
 & = & - z_{od} \, \partial_y \sum_1^K \delta_k \Delta_z
\end{eqnarray} 
where $\delta$ is the decrement from unity of the material's refractive index, $\beta$ its imaginary part and $\Phi_k=-(2\pi/\lambda) \delta_k \Delta_z$ is the phase shift introduced by the $k-$th object's layer. If the section thickness $\Delta y$ is small enough to allow a transition to the continuous formulation, we obtain  
\begin{eqnarray}\label{eq:att_tot_multislice_cont}
t  & = & e^{- (4 \pi / \lambda) \int \beta(z) dz } \\ \label{eq:ref_tot_multislice_cont}
\Delta y_R & = & - z_{od} \, \partial_y  \int \delta(z) dz
\end{eqnarray}
and similarly for the dark-field signal 
\begin{eqnarray}\label{eq:sca_tot_multislice_cont}
\sigma_O^2 & = & \int \sigma^2(z) dz.
\end{eqnarray}
Along with absorption and phase, also the dark-field signal $\sigma_O^2$ can be cast as an integral along the beam path. By collecting a number of different views while rotating the sample around the $y$ axis, the three dimensional distribution of the width of the object function $\sigma^2(x,y,z)$ can be calculated by means of, for example, the Inverse Radon Transform or the Filtered Back Projection algorithms as is routinely done for absorption and phase-based computed tomography. Equivalent results were obtained in the context of analyser based imaging \cite{Rigon:08_EurRad}, grating interferometry \cite{Bech:10_PMB} and beam-tracking \cite{Vittoria:15_SciRep}.

The experiment was performed at the SYRMEP beamline (Elettra Sincrotrone Trieste, Italy). The pre-sample slit was $40$ mm along $x$ and its aperture along $y$ was set at $20$ $\mu$m. The detector apertures were $23$ $\mu$m and arranged at a regular period of $79$ $\mu$m. The beam energy was $20$ keV, the pre-sample aperture was at about $22$ m from the source, the sample stage was at $26$ cm from the aperture and the analiser and detector a further $2.5$ m downstream. The detector was a CCD with GadOx scintillation screen (Photonic Science, UK) with a pixel size of 12.5 $\mu$m. 
 
For this experiment, five terms were retained for the illumination function (N=5) and the scattering distribution was assumed to be a single Gaussian (M=1). The three parameters representing absorption, refraction and scattering in the sample were obtained by a pixel-wise non-linear fitting procedure \cite{Endrizzi:15_APL} that compared the intensities recorded with and without the sample in the beam. For image acquisition, the illumination function was sampled in five positions with $\{ \pm 24, \pm 12, 0 \}$ $\mu$m displacement with respect to the position of maximum intensity in the central slit of the analyser, with an exposure time of $400$ milliseconds each.

The phantom used for planar imaging was composed of a cylinder of acrylic material with density $1.18$ g/cm$^3$ and radius $1.45$ mm and a step wedge made of paper layers. A melamine sponge prism was used to experimentally measure the dependency of the dark-field signal upon thickness. Mono disperse borosilicate micro spheres with diameters of $5$, $10$ and $12$ $\mu$m were embedded into an acrylic support for computed tomography acquisitions. A plastic scaffold was used to test the three-dimensional reconstruction on a phantom with a more complex geometry.

The data were recorded at a pixel size of $12.5$ $\mu$m and subsequently binned in such a way that the integrated intensity going through a single aperture in the detector mask was combined in a single image line. For the radiography image mode, the sample was scanned along $y$ in $15$ $\mu$m steps to build up a two dimensional image. For the computed tomography image mode, $600$ views were acquired with $0.3$ degree angular step. 

\begin{figure*}
    \centering
        \includegraphics[width=\textwidth]{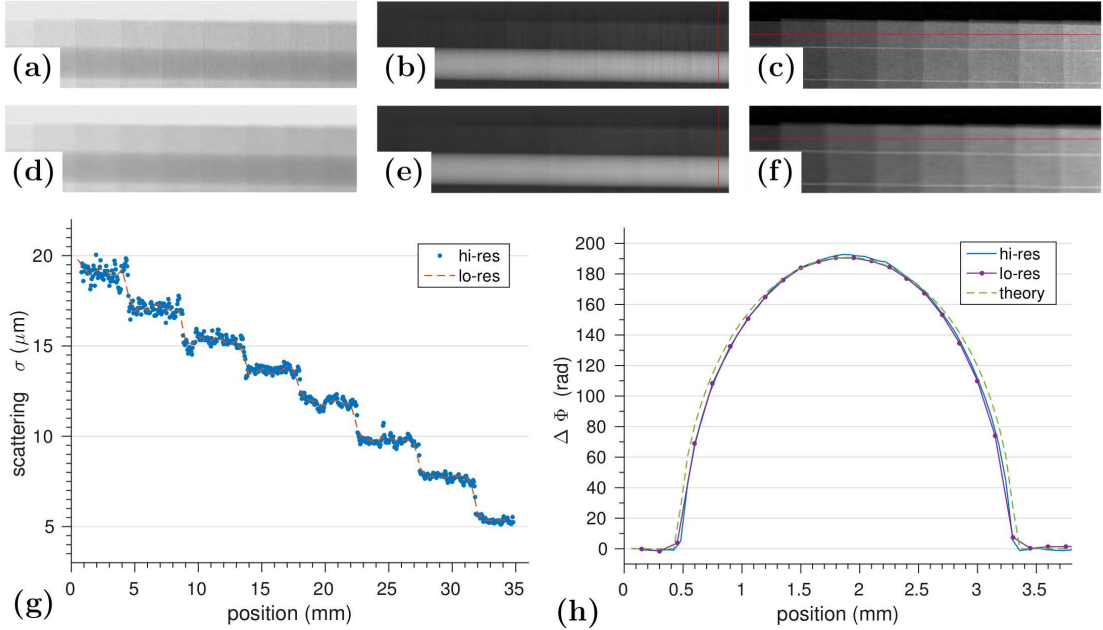}
\vspace{\baselineskip}
    \caption{Custom phantom consisting of a perspex cylinder and a paper step wedge: transmission (a) and (d), phase (b) and (e) and dark-field (c) and (f) images obtained by applying the retrieval to high resolution data. Panels (a)-(c) can be compared to the images obtained using the low resolution data, shown in panels (d)-(f). Quantitative profile plots reporting (g) the dark-field signal and (h) the phase shift measured along the lines highlighted in the panels (c),(f) and (b),(e) respectively, show that the spatial resolution does not affect the quantitativeness of the retrieved signals. A very good match betweeen theoretically expected and the experimentally measured profiles can be observed for the phase image.}\label{fig:ref_scat_sample}
\end{figure*}

The illumination function, acquired with a fine sampling step, is compared to the one obtained through a wave optics numerical simulation \cite{Vittoria:13_AO} that incorporates all the experimental parameters. A good match between the intensities experimentally recorded in the positions corresponding to the five apertures at the detector mask and the ones expected from the simulation can be observed in Figure \ref{fig:1}(b). 

This illumination function is taken as the reference one, and subsequently used for the retrieval of the sample images. The approximate rocking curve of a Si (111) crystal at the energy of $20$ keV is also plotted in Figure \ref{fig:1}(b), for a visual comparison of the sensitivity curves of the EI and the analyser crystal based imaging methods. It is interesting to note that the EI configuration reported here offers comparable curve width to that of the crystal analyser technique, at least when the $(111)$ diffraction is exploited in the latter.

\begin{figure*}
    \centering 
        \includegraphics[width=\textwidth, angle = 0]{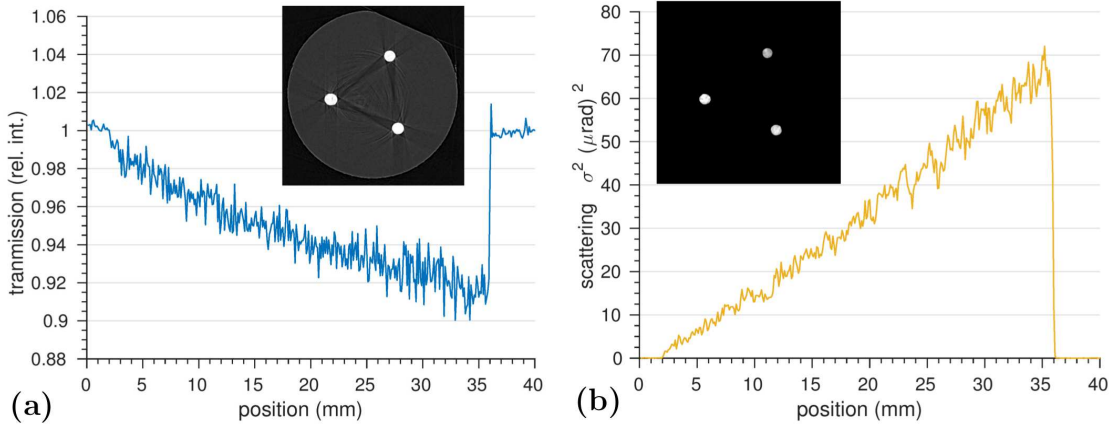}
    \caption{Quantitative profile plots reporting (a) transmission and (b) dark-field extracted from corresponding images of the melamine sponge prism sample. The dark-field signal $\sigma^2$ is directly proportional to the sample thickness, in accordance with the prediction of Equation \ref{eq:sca_tot_multislice_cont}. The two insets show computed tomography absorption (a) and dark-field (b) images of the microspheres embedded in the acrylic support.}\label{fig:scatVSthick}
\end{figure*}

In order to investigate the dependence of the retrieved signals on the spatial resolution, data were analysed two times. In the first case (high resolution) the pixel size was $50$ $\mu$m $\times$ $60$ $\mu$m while in the second case (low resolution) the pixel size was $500$ $\mu$m $\times$ $150$ $\mu$m. These two sets of data were obtained by binning the raw intensity acquisitions, after which the same analysis procedure is applied to both datasets. The results of this analysis are summarised in Figure \ref{fig:ref_scat_sample}. The effect of lowering the spatial resolution can be observed by comparing panels (a)-(c) to panels (d)-(f) in Figure \ref{fig:ref_scat_sample}, where the high- and low-resolution images are reported, respectively. For a quantitative comparison, line profiles are shown in the (g) and (h) panels of Figure \ref{fig:ref_scat_sample}. They were extracted along the directions highlighted in the panels (c), (f) for the plots in (g) and in the panels (b), (e) for the plots in (h). They represent the dark-field signal measured along the paper step wedge and the phase signal measured across the plastic cylinder. As it can be seen, the values obtained in the high- and low-resolution configurations are quantitatively very close to each other. Moreover, the theoretical phase shift shows a very good match with the experimentally measured one ($\delta = 6.61 \cdot 10^{-7}$ and $\beta = 3.34 \times 10^{-10}$ \cite{ts_imaging}), even though the experimental profiles in Figure \ref{fig:ref_scat_sample}(h) were extracted from the portion of the cylinder placed behind the part of the wedge with strongest scattering. The dispersion of the retrieved refraction angles, measured in an empty background region (around zero), can be used to estimate the sensitivity of the imaging system \cite{Modregger:11_OE, Diemoz:13_PRL, Diemoz:14_APL}. By following this procedure, the high and low spatial resolution configurations gave a standard deviation of $17$ nrad and $5.9$ nrad, respectively. The behaviour of transmission and dark-field signal as a function of thickness were investigated by means of the melamine sponge prism. Two line plots are reported in panels (a) and (b) of Figure \ref{fig:scatVSthick}. It can be clearly seen that the dark-field signal grows linearly with the sample thickness, and that the expected exponential relationship holds for the transmission. The dark-field CT result for the micro spheres sample is shown in the inset of Figure \ref{fig:scatVSthick}(b), along with the standard absorption CT image again showed as an inset in Figure \ref{fig:scatVSthick}(a). The values $\{0.95 \pm 0.06$, $1.10 \pm 0.05$, $1.19 \pm 0.08 \}$ $\cdot 10^3$ $\mu$rad$^2$ mm$^{-1}$ were measured for the $\{ 5$, $10$, $12 \}$ $\mu$m diameter spheres details, respectively. 
\begin{figure}
    \centering 
        \includegraphics[width=0.475\textwidth, angle = 0, trim = 40 50 30 60, clip]{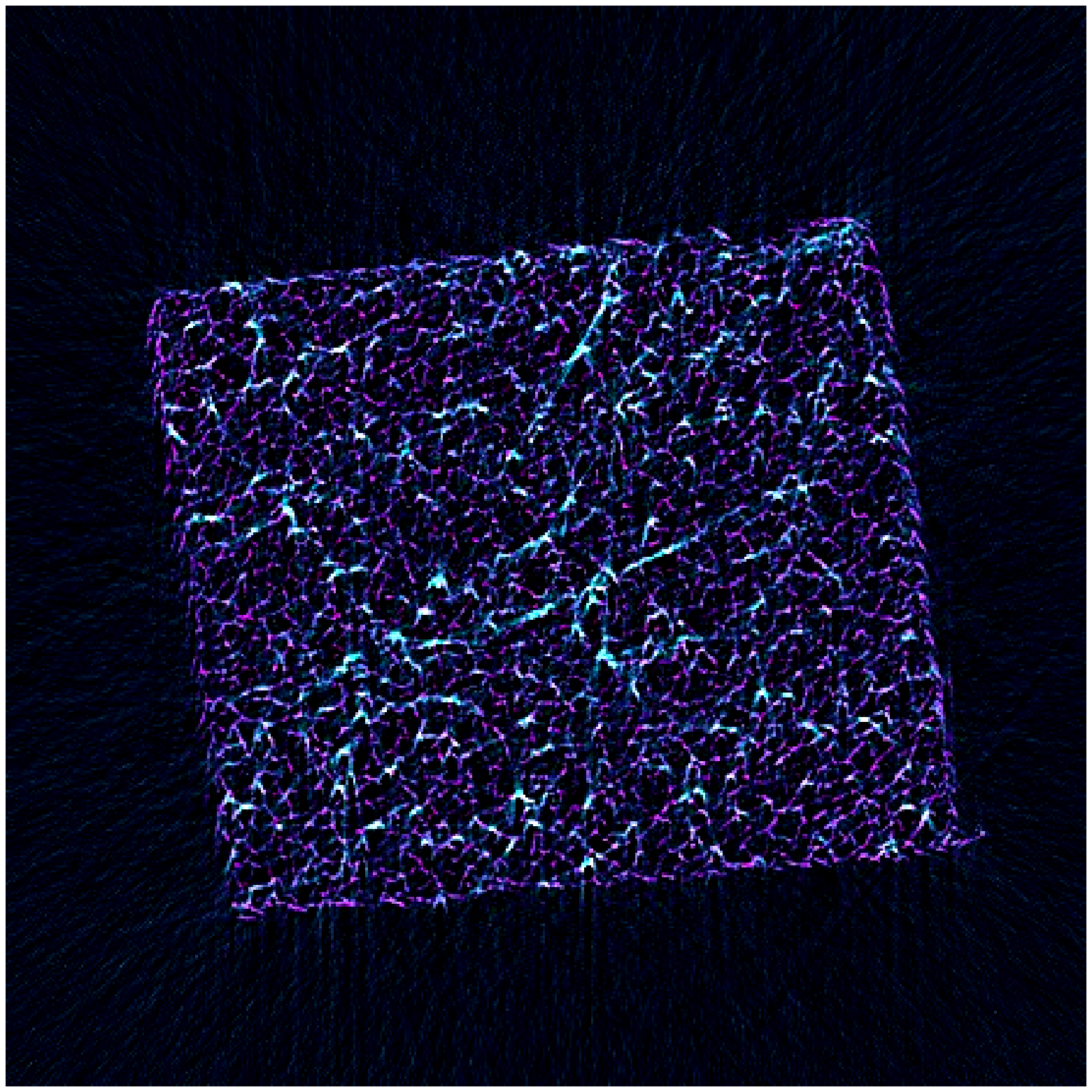}
\vspace{\baselineskip}
    \caption{Absorption and dark-field CT of a scaffold structure in false colours (absorption: magenta and dark-field: cyan), blended. It can be seen how the two different contrast channels provide a complementary representation of this complex sample.}\label{fig:scaff_blend}
\end{figure}
Finally, as an example for a phantom with a more complex geometry, the image obtained by fusing absorption- and dark-field- CT reconstructions of a plastic scaffold is shown in Figure \ref{fig:scaff_blend}. The two colour channels emerge from different sample details, meaning that they represent different properties of the sample and can offer a complementary visualisation of its characteristics.

In conclusion, we have presented a multi-aperture analiser set-up for radiography and computed tomography X-ray phase-contrast imaging. The linearity of the dark-field signal with the sample thickness was experimentally verified and quantitative computed tomography was performed on a custom phantom made of micro spheres embedded into an acrylic support. The method is based on the edge illumination method and it simultaneously provides absorption, refraction and dark-field contrast images. Very good agreement was observed between theory and experiment for the phase-shift image of a test sample. Notably, the phase shift was quantitatively measured to a high degree of accuracy also in the presence of strong scatterers in the beam. Finally, as an example of a complex geometry sample, the images of a scaffold were presented.

\section*{Acknowledgements}
This project was supported by the UK Engineering and Physical Sciences Research Council Grant EP/I021884/1. ME was supported by the Royal Academy of Engineering under the RAEng Research Fellowships scheme. We thank Elettra Sincrotrone Trieste for access to SYRMEP beamline (proposal 20140147) that contributed to the results presented here.

\bibliographystyle{apsrev4-1}

\end{document}